\documentclass{nature}
\usepackage{graphicx}
\makeatletter
\let\saved@includegraphics\includegraphics
\AtBeginDocument{\let\includegraphics\saved@includegraphics}
\renewenvironment*{figure}{\@float{figure}}{\end@float}
\makeatother
\usepackage{epstopdf}
\usepackage{dcolumn}
\usepackage{bm}
\usepackage{amssymb}
\usepackage{color}
\usepackage{amsmath}
\usepackage{soul}
\usepackage{float}
\usepackage[normalem]{ulem}

\usepackage{cancel}

\usepackage[colorlinks=true]
{hyperref}

\hypersetup{colorlinks,
citecolor=blue,
linkcolor=blue,
urlcolor=blue}
\title{Ultrafast optical rotation for extremely sensitive enantio-discrimination}

\author{David Ayuso$^{1,2}$, Andres Ordonez$^{1}$, Misha Ivanov$^{1,2,3}$ and Olga Smirnova$^{1,4}$} 

\begin{document}
\maketitle

\begin{affiliations}
\item Max-Born-Institut, Berlin, Germany
\item Department of Physics, Imperial College London, UK
\item Institute f\"ur Physik, Humboldt-Universit\"at zu Berlin, Germany
\item Technische Universit\"at Berlin, Germany
\end{affiliations}

\begin{abstract}
Sculpting sub-cycle temporal structures of optical waveforms allows one to image and even control electronic clouds in atoms\cite{Paulus2001Nat,Baltuska2003Nat}, molecules\cite{Kling2006Sci} and solids\cite{Schiffrin2013Nat,Luu2015}.
Here we show how the transverse spin component arising upon spatial confinement of such optical waveforms enables extremely efficient chiral recognition and control of ultrafast chiral dynamics.
When an intense few-cycle, linearly polarized laser pulse is tightly focused into a medium of randomly oriented chiral molecules, the  medium generates light which 
is elliptically polarized, with opposite helicities and opposite rotations of the polarization ellipse in media of opposite handedness.
In contrast to conventional optical activity of chiral media, this new nonlinear optical activity is driven by purely electric-dipole interactions and leads to giant enantio-sensitivity in the near VIS-UV domain, where optical instrumentation is readily available.
Adding a polarizer turns rotation of the polarization ellipse into highly enantio-sensitive intensity of the nonlinear-optical response.
Sub-cycle optical control of the incident light wave enables full control over the enantio-sensitive response. 
The proposed all-optical method not only enables extremely efficient chiral discrimination, but also ultrafast imaging and control of chiral dynamics with commercially available optical technology.
\end{abstract}

Chirality plays key role in nature, both at microscopic and macroscopic levels, from controlling the behavior of sub-atomic particles\cite{book_Kane1993} and the biological activity of drugs\cite{book_Berova2013} to shaping galaxies\cite{Kondepudi2001}.
Chiral molecules can exist in two mirror-reflected versions, known as the left- and right-handed enantiomers.
These mirror twins have identical physical properties and identical chemical reactivity unless they interact with another chiral object, such as circularly polarized light, which  has long served as a standard tool for chiral recognition\cite{book_Berova2013}.
However, circularly polarized light
is not efficient: the chiral structure of the light helix emerges only beyond of the electric-dipole approximation.
The weakness of non-dipole interactions makes the optical response of left- and right-handed media to this field virtually identical, and chiral discrimination hard.
One way around this fundamental limitation is to use \emph{synthetic} chiral light\cite{Ayuso2019NatPhot}, which is chiral already within the electric-dipole approximation and can lead to huge enantio-sensitivity.

However, it has also been known since the time of Luis Pasteur\cite{Pasteur1848} that the chirality of light is not required to detect the chirality of matter.
Even linearly polarized light, which is achiral, leads to optical activity: when propagating through a chiral medium, its polarization plane rotates in opposite directions in media of opposite handedness.
While this effect relies on weak non-dipole interactions, it benefits from the cumulative addition from all chiral units in optically dense media, eventually leading to large rotation angles.
Still, detecting the handedness of dilute samples, such as those routinely produced in standard chemistry labs, is very challenging.
Several techniques\cite{Ritchie1976PRA,Powis2000JCP,Bowering2001PRL,Lux2012Angew,Stefan2013JCP,Kastner2016CPC,Beaulieu2018NatPhys,Goetz2019PRL,Demekhin2018PRL,RozenPRX2019,Fischer2000PRL,Ji2006JACS,Belkin2000PRL,Patterson2013Nat,Tutunnikov2018JPCL}
can yield strongly enantio-sensitive signals in thin and low-concentration media by analyzing vectorial observables\cite{OrdonezPRA2018}, 
such as the direction of the photo-electron current\cite{Ritchie1976PRA,Powis2000JCP,Bowering2001PRL,Lux2012Angew,Stefan2013JCP,Kastner2016CPC,Beaulieu2018NatPhys} upon ionization with circularly polarized light, where the chiral response arises due to purely electric-dipole interactions.
The application of two-colour cross-polarized beams and quantum control strategies may further increase the enantio-sensitivity of this method\cite{Demekhin2018PRL,Goetz2019PRL,RozenPRX2019}.
Even enantio-sensitive attosecond delays in photo-ionization have recently been reported\cite{Beaulieu2017Science}.
However, these signals require angle-resolved detection of photo-electrons, rather than purely optical signals, which restricts their applicability.

Here we show how intense linearly polarized few-cycle laser pulses, when confined both in time and in space, can generate an extremely strong nonlinear analogue of optical activity, enabling efficient chiral discrimination of dilute samples with commercially available optical technology.
Intense few-cycle laser pulses have opened a range of new opportunities for ultrafast spectroscopy\cite{Krausz2009RMP}, both in atoms\cite{Paulus2001Nat,Baltuska2003Nat} and solids\cite{Schiffrin2013Nat,Luu2015}.
When the electric field amplitude $E_0$ of the laser pulse $E(t)=E_0a(t)\cos(\omega t+\phi_{\text{CEP}})$ is strongly modulated by the rapidly varying envelope $a(t)$,
the carrier-envelope phase (CEP) $\phi_{\text{CEP}}$ determines the temporal structure of the electric field oscillations, and thus the nonlinear response of matter.
In the frequency domain, the large coherent bandwidth of ultrashort pulses gives rise to interference between different-order multiphoton pathways.
This interference is sensitive to the spectral phase in general, and to $\phi_{\text{CEP}}$ in particular, giving rise to $\phi_{\text{CEP}}$-dependent low-order \cite{Mehendale2000} and high-order harmonic generation \cite{Haworth2007NatPhys} and photo-electron emission\cite{Cormier1998,Dietrich2000,Paulus2001Nat,Wittmann2009NatPhys}, which encode $\phi_{\text{CEP}}$ with high precision.

Interferences of different multiphoton pathways also
open new  opportunities for enantio-sensitive imaging and control of ultrafast chiral dynamics in molecules.
Below we present a simple, all-optical setup that encodes the handedness of a chiral medium in the polarization properties of the nonlinear response.
We show that the emitted harmonic light exhibits optical activity: opposite ellipticity and opposite rotation angles in media of opposite handedness.
Adding a polarizer makes the detection setup chiral and turns enantio-sensitive polarizations into highly enantio-sensitive intensities of optical signals.

\begin{figure}[H]
\centering
\includegraphics[width=\textwidth, keepaspectratio=true]{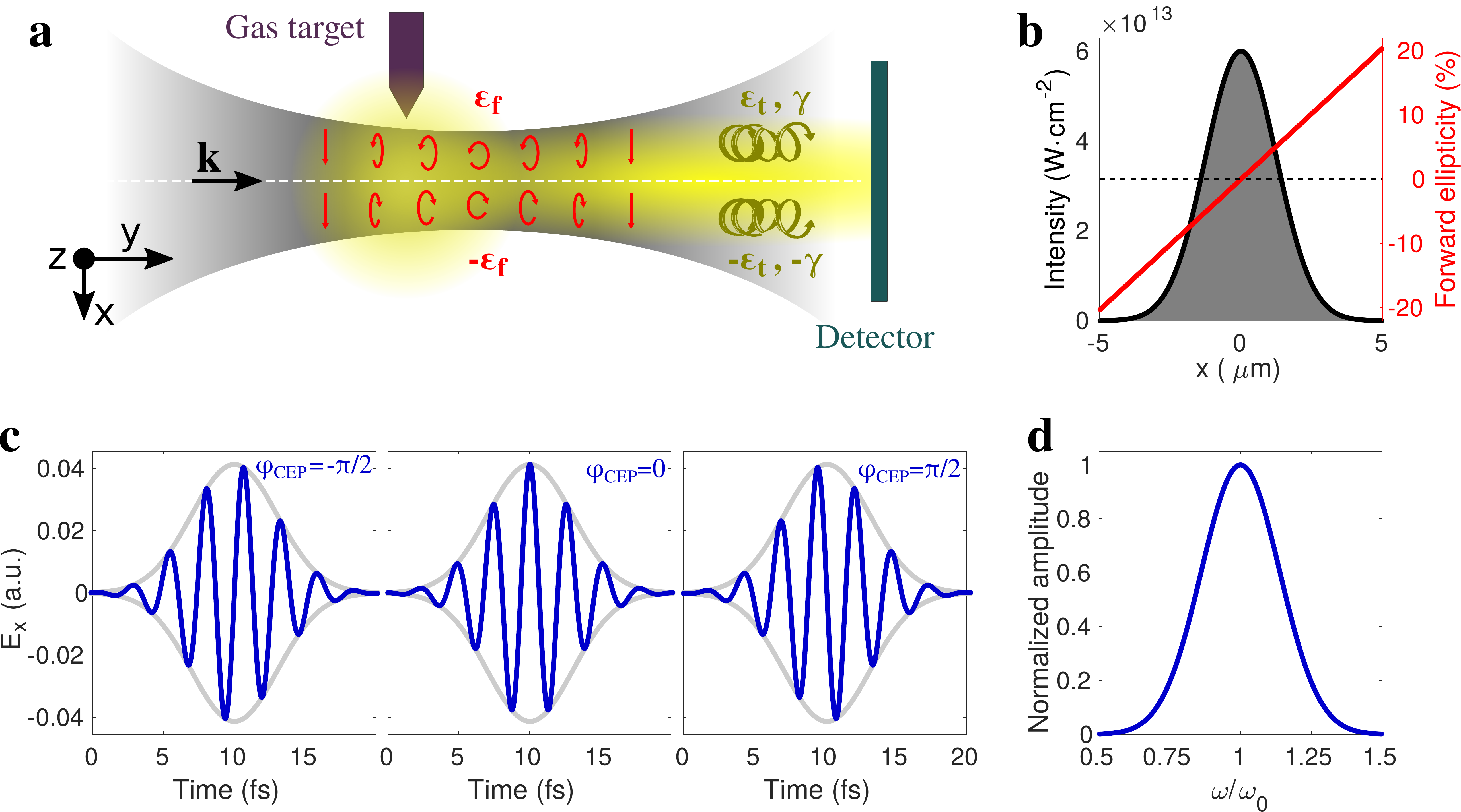}
\caption{\textbf{Proposed experimental setup.}
\textbf{a,} A few-cycle, $x$-polarized, tightly focused beam acquires ellipticity $\varepsilon_f$ along the propagation direction $y$, with opposite signs at opposite sides of the beam axis.
A gas jet of chiral molecules, placed before the focus, generates elliptically polarized harmonics.
The transverse ellipticity and rotation angle of the harmonic polarization ellipses $\varepsilon_t$ record the molecular handedness, and have opposite sign at the opposite sides of the beam.
\textbf{b,} Intensity (black) and forward ellipticity (red) of the driving field.
\textbf{c,d} Electric field in the time (\textbf{c}) and frequency (\textbf{d}) domains at the beam axis ($x=0$), for $\phi_{\text{CEP}}=-\pi/2, 0, \pi/2$ (the frequency-domain amplitude is CEP-independent).
}
\label{fig_setup}
\end{figure}

The proposed all-optical method relies on standard lasers to generate ultrashort pulses, which are then tightly focused into a gas jet of chiral molecules, see Figs. \ref{fig_setup}a,b.
Commercially available technology allows one to generate CEP-stable pulses lasting only a few cycles (confinement in time, Fig. \ref{fig_setup}c), with broad spectral bandwidth (Fig. \ref{fig_setup}d).
By focusing the beam tightly (confinement in space, Figs. \ref{fig_setup}a,b), we create a strong longitudinal field\cite{Bliokh2015}
\begin{equation}\label{eq_longitudinal}
E_y = \pm \Re \bigg\{ \frac{i}{k} \frac{\partial E_x}{\partial x} \bigg\}
\end{equation}
where $E_x$ is the transverse field component (see Fig. \ref{fig_setup}a), $k=\frac{2\pi}{\lambda}$ is the wave number, $\lambda$ is the wavelength. The symbol $\pm$ is $+$ when the beam propagates along $z$ and $-$ when it propagates along $-z$.
As a result of spatial confinement, the field acquires forward ellipticity $\varepsilon_f=E_y/E_x$, i.e. ellipticity in the propagation direction.
For a linearly polarized Gaussian beam, the forward ellipticity is
\begin{equation}
\varepsilon_f = \pm \frac{2x}{k\tilde{w}^2}
\end{equation}
where $\tilde{w}$  is the beam waist (see Fig. \ref{fig_setup}b and Methods).
The sign of the forward ellipticity, or the transverse spin component, is opposite on opposite sides of beam axis, and is locked to the direction of light propagation.
This direct mapping between transverse spin and propagation direction, the so-called spin-momentum locking\cite{Lodahl2017Nat}, also appears naturally in optical nanofibers and other nano-photonic structures, where it gives rise to extraordinary phenomena, such as propagation-direction-dependent emission\cite{Mitsch2014NatComm,Sollner2015NatNano}.
Here we exploit it for highly efficient laser-controlled chiral discrimination with few-cycle pulses.

Consider first the nonlinear response of isotropic chiral media to such light.
As both the field and the medium have mirror symmetry with respect to the $xy$ plane (Fig. \ref{fig_setup}), the induced polarization is confined to this plane. Since the $y$-polarization component is parallel to the propagation direction, it does not contribute to the far-field signal. Thus, the far-field harmonic emission is linearly polarized along $x$.
In contrast, in isotropic chiral media, where the mirror symmetry is absent, polarization along $z$ is no longer forbidden, and has opposite phase in media of opposite handedness.
This chiral polarization component can be strong, because it is driven by purely electric dipole interactions.

For long pulses, symmetry dictates that the chiral ($z$-polarized) and achiral ($x$-polarized) components have different parity, appearing at even and odd harmonics of the fundamental frequency, respectively.
However, this does not apply to broadband few-cycle pulses, where the $\phi_{\text{CEP}}$-dependent interference of even- and odd-order multiphoton pathways is the norm, and the orthogonal chiral and achiral polarization components overlap in frequency.
Moreover, the $z$-polarized chiral response is out of phase in opposite enantiomers. Consequently, the molecular handedness is mapped on opposite rotations and opposite circularities of the polarization ellipse  of the generated light.
Control of $\phi_{\text{CEP}}$ enables full control of this enantio-sensitive response.

\begin{figure}[H]
\centering
\includegraphics[width=\textwidth, keepaspectratio=true]{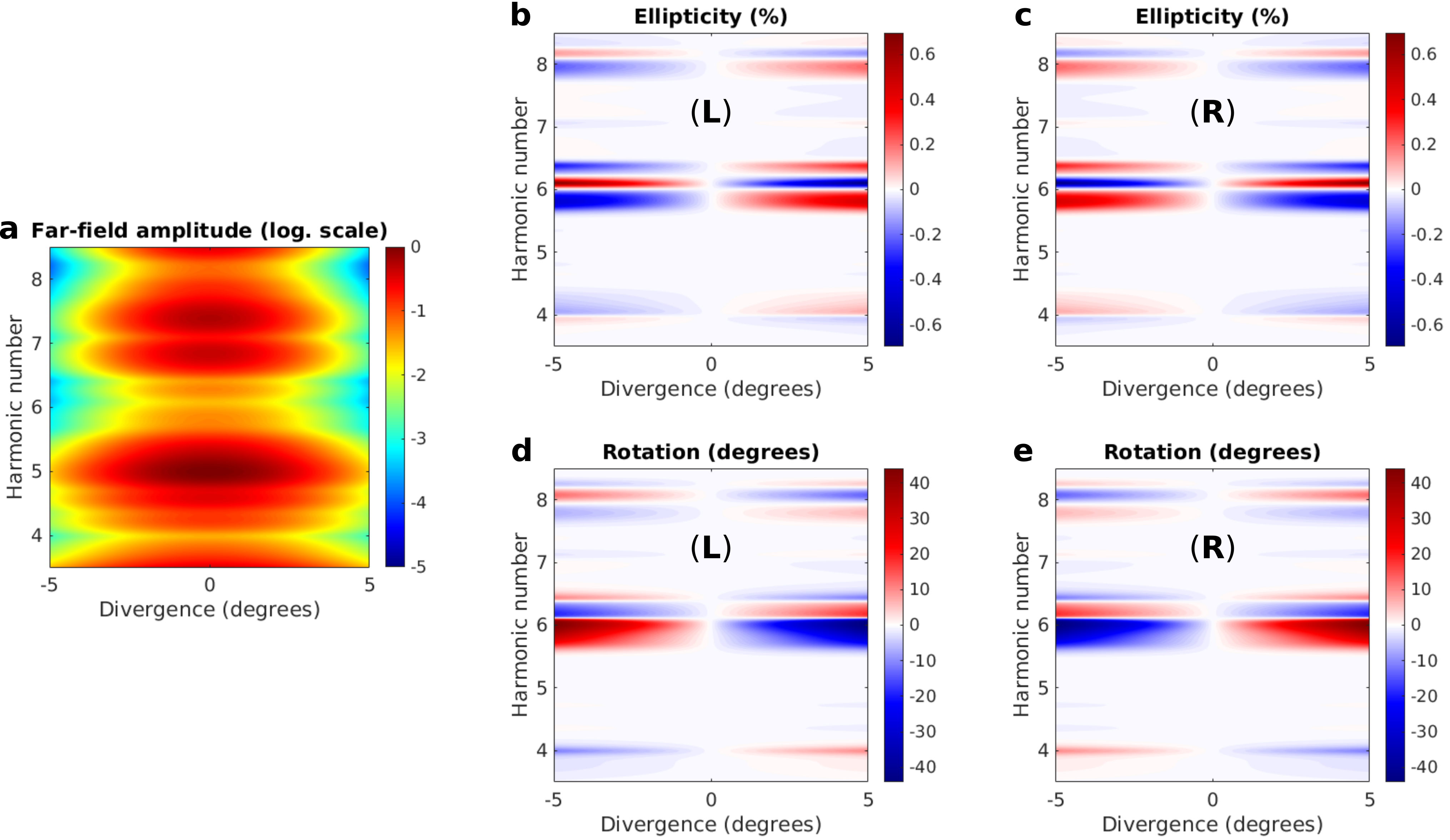}
\caption{\textbf{Ultrafast and nonlinear optical rotation.}
\textbf{a,} Far-field amplitude of the harmonic light emitted by randomly oriented propylene oxide molecules 
driven by the field shown in Fig. \ref{fig_setup},
versus the harmonic number and the emission angle.
The laser parameters are: intensity $6\cdot10^{13}$ W$\cdot$cm$^{-2}$, $\phi_{\text{CEP}}=\pi/2$, focal diameter $5 \mu$m,  carrier wavelength $\lambda=780$nm, and pulse duration $7$fs (full width at half maximum of the field's amplitude).
\textbf{b-e,} Ellipticity (\textbf{b,c}) and rotation angle (\textbf{d,e}) of the polarization ellipses of the harmonic light emitted from left-handed (\textbf{b,d}) and right-handed (\textbf{c,e}) molecules.}
\label{fig_ff}
\end{figure}

We have calculated the nonlinear response of randomly oriented propylene oxide molecules to the field in Fig. \ref{fig_setup} using  a state-of-the-art implementation of time-dependent density functional theory, see Methods.
The harmonic field emitted by right- and left-handed molecules can be written as $\mathbf{F}_{R,L} = F_{a} \mathbf{x} \pm  F_{c} \mathbf{z}$, where ``$+$" stands for right-handed and ``$-$" stands for left-handed molecules.
The intensity and polarization of the emitted harmonic light in the far field are shown in Fig. \ref{fig_ff}, for $\phi_{\text{CEP}}=0.5\pi$ (see Fig. \ref{fig_setup}c).
The total harmonic intensity (Fig. \ref{fig_ff}a), proportional to $|F_{a}|^2+|F_{c}|^2$, is not enantio-sensitive because the incident light is achiral.
However, the molecular handedness controls the polarization of the emitted harmonics.
Figs. \ref{fig_ff}b-e show the ellipticity and rotation angle of the polarization ellipses of light emitted by left- and right-handed molecules.
Both quantities reach very large values in wide spatial and spectral regions, allowing for 
accurate detection of the medium handedness using standard optical instrumentation.
Giant enantio-sensitivity is observed near harmonics 4, 6 and 8, corresponding to wavelengths of $\lambda=195, 130$, and $97.5$nm, respectively.
As the transverse spin of the driving field has opposite sign at opposite sides of the beam axis (Figs. \ref{fig_setup}a,b), so do the harmonic ellipticity and the polarization rotation angle. 

This new type of optical activity opens a way for highly efficient detection of chiral matter and ultrafast imaging of chiral dynamics.
It also enables accurate determination of the enantiomeric excess in macroscopic mixtures, $ee=\frac{C_R-C_L}{C_R+C_L}$, where $C_R$ and $C_L$ are the concentrations of right- and left-handed molecules, respectively.
Both the ellipticity and the rotation angle depend linearly on the enantiomeric excess (see Methods), $\varepsilon(ee)=ee\cdot\varepsilon_R$ and $\gamma(ee)=ee\cdot\gamma_R$, where $\varepsilon_R$ and $\gamma_R$ are the ellipticity and the rotation angle recorded in an enantio-pure sample of right-handed molecules.

\begin{figure}[H]
\centering
\includegraphics[width=14cm, keepaspectratio=true]{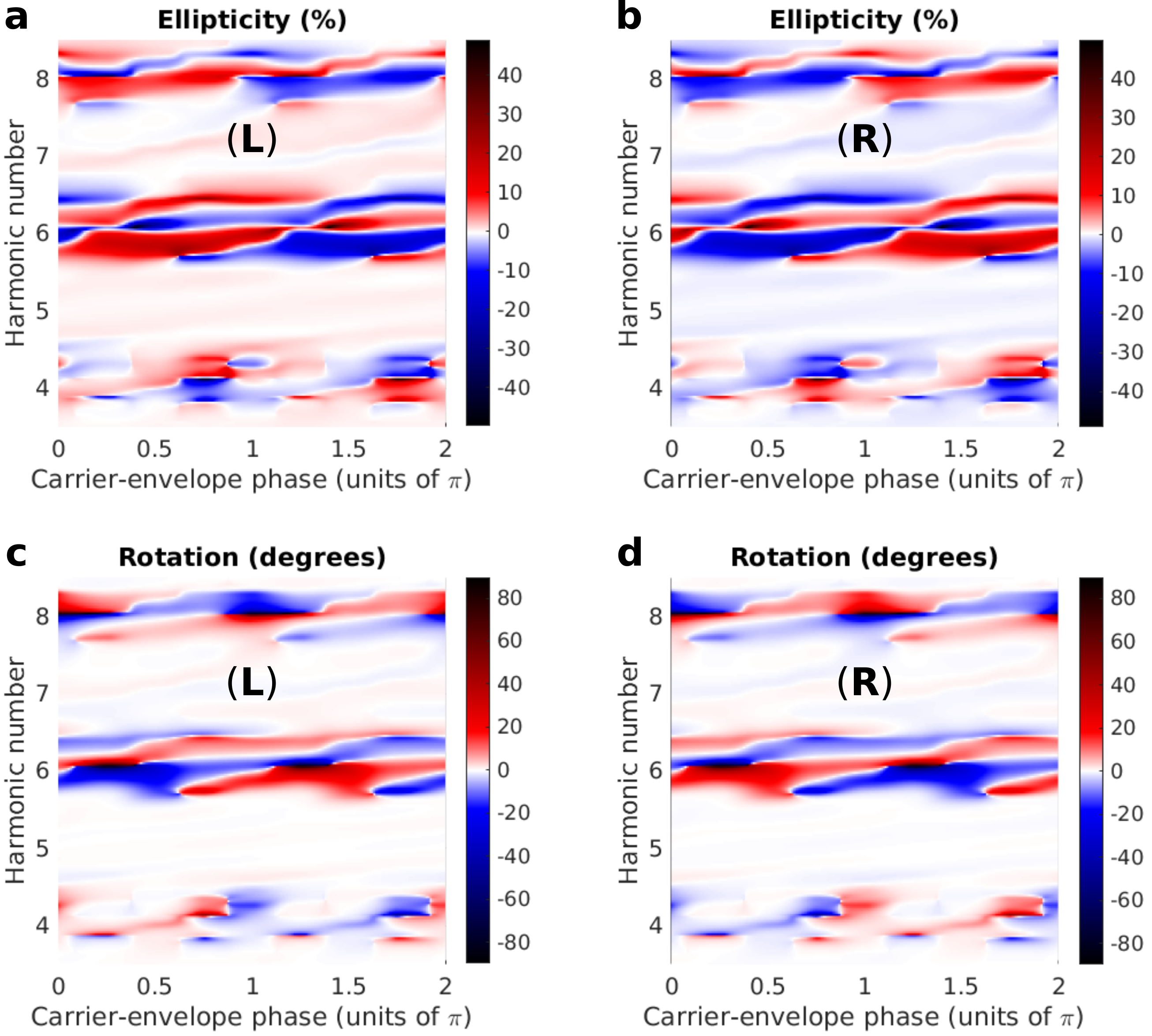}
\caption{\textbf{CEP-control over the enantio-sensitive polarization.}
Ellipticity (\textbf{a,b}) and rotation angle (\textbf{c,d}) of the polarization ellipses of the harmonic light emitted from the left-handed (\textbf{a,c}) and right-handed (\textbf{b,d}) propylene oxide molecules at emission angle $\beta=3^{\circ}$, as functions of the harmonic number and the CEP.
Laser parameters are the same as in Fig.2.}
\label{fig_ff2}
\end{figure}

The polarization (ellipticity and rotation angle) of the emitted harmonic light records the relative amplitudes and phases of the chiral ($z$-polarized) and achiral ($x$-polarized) components of the light-driven dynamics.
Shaping the sub-cycle temporal structure of the incident light wave allows us to fully control it.
Fig. \ref{fig_ff2} shows that the ellipticity and rotation angle of the emitted light can be controlled by the CEP of the driving field.
We present the values recorded at a fixed emission angle $\beta=3^{\circ}$ --note that the results for other angles are similar, see in Figs. \ref{fig_ff}b-e.
The direct mapping between the light-driven chiral dynamics and the polarization of harmonic light allows one to reconstruct the ultrafast electronic response of a chiral molecule.
Specifically, measuring the intensity, ellipticity and rotation angle of the emitted harmonic light allows us to reconstruct both the amplitude and phase of the ultrafast chiral response, see Methods.

\begin{figure}[H]
\centering
\includegraphics[width=\textwidth, keepaspectratio=true]{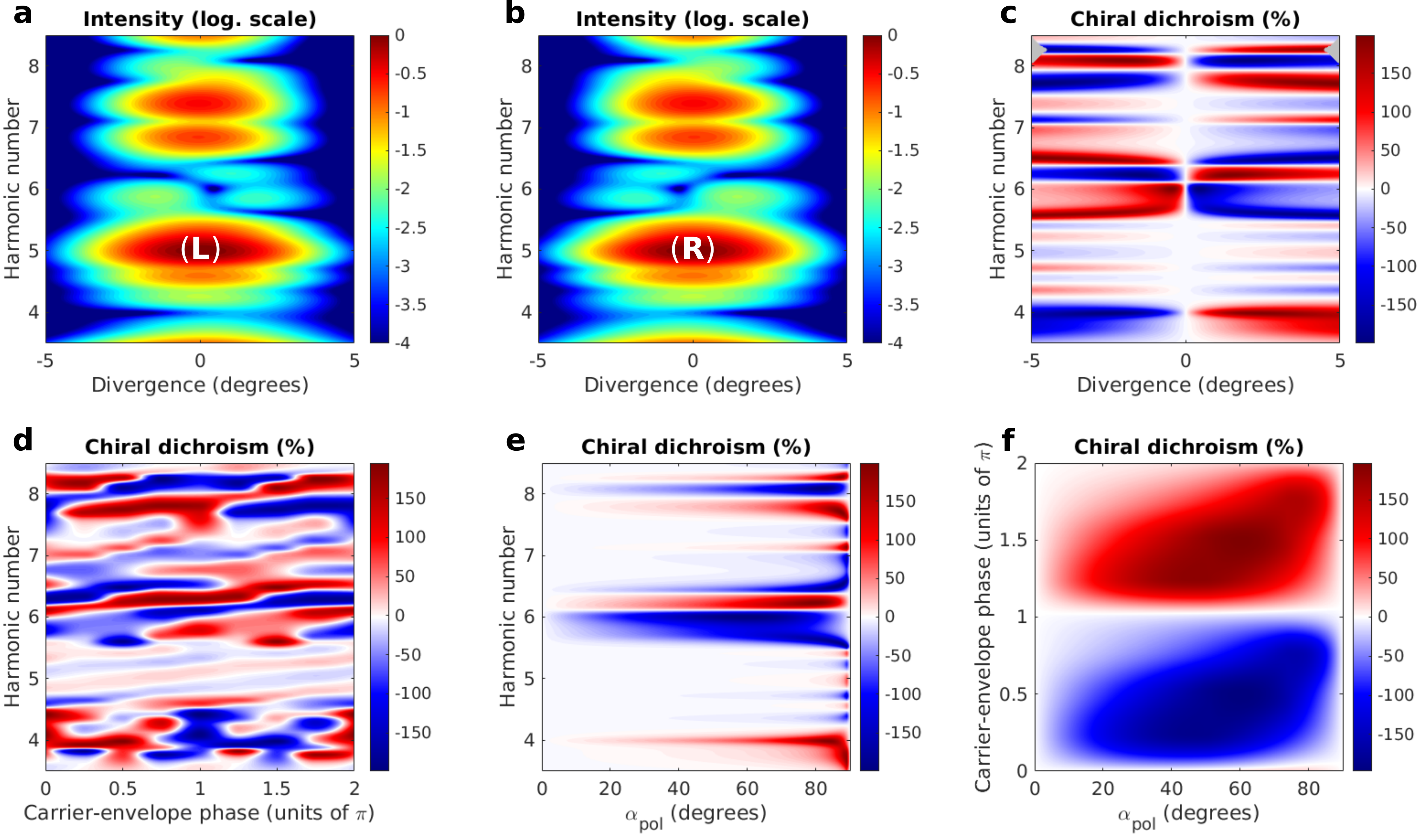}
\caption{\textbf{Turning enantio-sensitive polarization into enantio-sensitive intensity.}
\textbf{a,b}, Harmonic intensity from left-handed (\textbf{a}) and right-handed (\textbf{b}) propylene oxide driven by the setup of Fig. \ref{fig_setup} for $\phi_{\text{CEP}}=\pi/2$, when using a polarizer with $\alpha_{\textrm{pol}}=85^{\circ}$ (see main text).
\textbf{c-f}, Chiral dichroism in the harmonic intensity, $CD=2\frac{I_L-I_R}{I_L+I_R}$, for
fixed $\phi_{\text{CEP}}=\pi/2$ and $\alpha_{\textrm{pol}}=85^{\circ}$ (\textbf{c}),
fixed $\alpha_{\textrm{pol}}=85^{\circ}$ and $\beta=3^{\circ}$ (\textbf{d}),
fixed $\phi_{\text{CEP}}=\pi/2$ and $\beta=3^{\circ}$ (\textbf{e}), and
fixed harmonic number $N=6$ and $\beta=3^{\circ}$ (\textbf{f}).
The other laser parameters are the same as in Fig.2.}
\label{fig_pol}
\end{figure}

The highly enantio-sensitive polarization of the emitted harmonic light can be converted into highly enantio-sensitive intensities.
Placing a polarizer before the detector, we can project both achiral ($F_{a}$) and chiral ($\pm F_{ch}$) components of the harmonic radiation onto the same polarization axis, and make them interfere.
The harmonic intensity becomes proportional to $I_{R,L}(\alpha_{\text{pol}}) \propto \big| \cos(\alpha_{\text{pol}}) F_{a} \pm \sin(\alpha_{\text{pol}}) F_{ch} \big|^2$,
where $\alpha_{\text{pol}}$ is the angle between the selected polarization direction and the $x$ axis.
Figs. \ref{fig_pol}a,b show the harmonic intensity emitted from left- and right-handed propylene oxide molecules at the detector, for $\alpha_{\text{pol}}=85^{\circ}$.
The controlled interference between $F_{a}$ and $F_{ch}$ makes the harmonic intensity strongly enantio-sensitive, and the chiral dichroism, $CD=2\frac{I_L-I_R}{I_L+I_R}$, reaches the ultimate efficiency limit of $200\%$, as shown in Figs. \ref{fig_pol}c-f.
The possibility of measuring enantio-sensitive intensities using achiral driving field (Fig. \ref{fig_setup}) stems from the fact that this field, together with the polarizer, create a chiral measurement setup, realizing the concept of a ``chiral observer" \cite{OrdonezPRA2018}.

By adjusting the angle of the polarizer $\alpha_{\text{pol}}$ and the CEP of the driver, we can achieve perfect constructive or destructive interference for any harmonic frequency.
Indeed, the relative strength between $F_{a}$ and $F_{ch}$ can be adjusted with $\alpha_{\text{pol}}$, and their relative phase is controlled by the CEP.
Fig. \ref{fig_pol}d shows the chiral dichroism as a function of the harmonic number and the CEP, for $\alpha_{\text{pol}}=85^{\circ}$ and emission angle $\beta=3^{\circ}$.
Note that the sign of the chiral dichroism is not affected by the value of $\alpha_{\text{pol}}$, as shown in Figs. \ref{fig_pol}e,f.

The chiral dichroism in the emitted harmonic light provides direct access to chiral electron dynamics.
In particular, the ``white lines'' in Fig. \ref{fig_pol}d indicate the CEP values for which the achiral ($x$-polarized) and chiral ($z$-polarized) components have a phase delay of $\pm\pi/2$, and thus the polarization vector draws an ellipse in the $xz$ plane at frequency $N\omega$ with its major axis along $x$ (or $y$).
On the other hand, the CEP values that maximize chiral dichroism correspond a phase delay of $0$ or $\pi$, so that the polarization vector at frequency $N\omega$ forms a straight line in the $xz$ plane.

The proposed method opens unique opportunities for efficient control and imaging of ultrafast chiral dynamics using commercially available optical technology.
The possibility of reconstructing the ultrafast electronic response of chiral molecules to light via multi-dimensional spectroscopy, with CEP and frequency being the two measurement dimensions, from either polarization or intensity measurements, can be exploited to visualize changes in molecular chirality during photo-induced chemical reactions, in real time. 
Adding more spectroscopic dimensions, e.g. a chirp parameter, may enhance these opportunities even further.

One of the key ingredients of the proposed new method is the longitudinal field component that naturally arises when light is confined in space.
Such longitudinal fields are not exclusive of tightly focused laser beams\cite{Bliokh2015}.
They appear in a number of physical environments, e.g. in optical nanofibers and periodic nano-photonic structures, where light is confined in one or two spatial dimensions\cite{Lodahl2017Nat}.
New opportunities may arise from creating controlled waveforms inside these nano-structures.
They may allows one to design compact devices for efficient chiral discrimination and imaging, to be used for routine operations in standard chemistry labs.

\section*{Methods}

\subsection{Modeling the driving field in the interaction region.}

We are interested in describing the interaction of a linearly polarized Gaussian beam with a gas jet of randomly oriented molecules which is placed before the beam's focus, as in standard HHG experiments.
If the medium is sufficiently thin, we can assume that the properties of the driving field remain constant along the propagation direction $z$ all over the interaction region.
Then, the transverse polarization component can be written as
\begin{equation}
E_x(x,y) = E_0 e^{-(x^2+y^2)/w^2} \cos(\omega t)
\end{equation}
Its longitudinal component (see Eq. \ref{eq_longitudinal}) is given by
\begin{equation}
E_y(x,y) = 2 E_0 \frac{x}{kw^2} e^{-(x^2+y^2)/w^2} \sin(\omega t)
\end{equation}
The weak spatial modulation of $E_x$ along $x$ in weekly-focused beams (with large $kw^2$) leads to vanishing $E_y$.
However, this longitudinal component can be strong when light is confined in space, as in the tightly focused beam considered here, with $w=2.5\cdot10^{-6}\mu$m, which creates strong forward ellipticity, see Fig. 1b.
We have evaluated the nonlinear response of propylene oxide as a function of the transverse coordinate $x$, as described in the following.

\subsection{Single-molecule response of propylene oxide.}

The ultrafast electronic response of the chiral molecule propylene oxide to the proposed light field has been evaluated in the framework of time-dependent density functional theory, using the state-of-the-art implementation in Octopus \cite{Tancogne2020,Andrade2015,Castro2006,Marques2003}.
Electronic exchange and correlation effects have been accounted for using the local-density approximation\cite{Dirac1930,Bloch1929,Perdew1981}.
We used the averaged-density self-interaction correction\cite{Legrand2002} to ensure the adequate description of the part of the electronic density that is driven far away from the molecule by the intense laser field.
The 1s orbitals of the carbon and oxygen atoms were described via pseudo-potentials, as they do not play a role in the physical processes that we tackle.
Octopus expands the Kohn-Sham orbitals and the electron density into a uniform real-space grid, allowing us to control the quality of the discretization with the spacing between neighboring points $\Delta x$.
We have used a spherical basis set of radius $R=42$ a.u. and $\Delta x=0.4$ a.u. to reach convergence, with absorbing boundary conditions to avoid unphysical reflexions of the electron density.
We used a complex absorbing potential with width $20$ a.u. and height $-0.2$ a.u.

We performed calculations for different molecular orientations in order to describe the physical situation of randomly oriented molecules.
The total polarization results from the coherent addition of the contribution from all possible orientations:
\begin{equation}
\mathbf{P}(t) = \int d\Omega \int d\alpha \, \mathbf{P}_{\Omega,\alpha}(t)
\end{equation}
The integration in the solid angle $\Omega$ was performed using a Lebedev quadrature\cite{Lebedev1999} and, for each value of $\Omega$, we integrated over $\alpha$ using the trapezoid method, as in our previous works\cite{Ayuso2018JPB,Ayuso2018JPB_model,Neufeld2019PRX}.
Here we used the Levedeb quadrature of order 11 (50 angular points in $\Omega$) and 4 points in $\alpha$.
The laser field was rotated in the molecular frame in a way that, for each value of $\Omega$, the strong field component always pointed along the same direction, which allowed us to reach convergence using a small number of points in $\alpha$.

The procedure above described was applied to evaluate the electronic response of randomly oriented propylene oxide to laser fields with $5\%$ of ellipticity, intensity $I=6\cdot10^{13}$W$\cdot$cm$^{-2}$ and varying CEP.
This intensity corresponds to the maximum value in the intensity profile, see Fig. \ref{fig_setup}b.
The single-molecule response in other points in the interaction region was calculated assuming that the nonlinear response along $x$, which is essentially driven by the strong-field component $F_x$, depends on the number of absorbed photons as $D_x \propto I^{N/2}$,
and that $D_z \propto I^{N/2}\varepsilon$.
The nonlinear response along $y$ is not observed in the far field as it is parallel to the direction of light propagation.

We have run numerical simulations for the right-handed enantiomer of propylene oxide, for four values of the CEP: $0$, $\pi/2$, $\pi$ and $3\pi/2$.
The results for CEP$=\pi$, $5\pi/2$, $3\pi$ and $7\pi/2$, and for the left-handed enantiomer, were obtained using symmetry considerations.
The CEP-dependent values of ellipticity, rotation angle and chiral dichorism presented in Figs. \ref{fig_ff2} and \ref{fig_pol}d were obtained using the Piecewise Cubic Hermite Interpolating Polynomial method implemented in Matlab.

\subsection{Enantio-sensitive polarization in the far field.}

For each frequency component, the chiral ($F_z$) and achiral ($F_x$) components of the emitted harmonic light in the far field were evaluated using the Fraunhofer diffraction equation:
\begin{align}\label{Fraunhofer}
F_x(\beta) \propto \int_{-\infty}^{\infty} \frac{d^2}{dt^2} P_x(x) e^{-iK x} dx \\
F_z(\beta) \propto \int_{-\infty}^{\infty} \frac{d^2}{dt^2} P_z(x) e^{-iK x} dx
\end{align}
where $P_x(x)$ and $P_z(x)$ and are achiral and chiral light-driven polarization components along the transverse coordinate $x$in the near field (see previous section), and $\beta$ is the far field angle (divergence),
with $K=\frac{N\omega}{c}\beta$ ($c$ is the speed of light in vacuum).
The ellipticity $\varepsilon$ and rotation angle $\gamma$ of each frequency component of the emitted light are realted to $F_x$ and $F_z$ via
\begin{align}
\tan(2\gamma) &= \tan(2\Psi) \cos(\delta) \label{eq_gamma} \\
\sin(2\chi)   &= \sin(2\Psi) \sin(\delta) \label{eq_chi}
\end{align}
where $\tan(\Psi)=|F_z|/|F_x|$, $\delta=\arg(F_z)-\arg(F_x)$ and $\varepsilon = \tan(\chi)$.

The relative concentration of opposite molecular enantiomers in a mixture is usually quantified using the enantiomeric excess, $ee=\frac{C_R-C_L}{C_R+C_L}$, where $C_R$ and $C_L$ are the concentrations of right- and left-handed molecules, respectively.
Let us analyze how the polarization of the emitted harmonic light depends on this quantity assuming, for simplicity, that $C_R+C_L$ remains constant.
The achiral component $F_x$ depends on the total number of molecules, not on their handedness, and thus it is $ee$-independent.
However, as the chiral component $F_z$ is emitted out of phase from opposite enantiomers, the amplitude of the z-polarized emission signal depends linearly on $ee$.
As a result (see Eqs. \ref{eq_gamma} and \ref{eq_chi}), both the $\varepsilon$ and $\gamma$ depend linearly on $ee$.

\section*{Acknowledgements}
We thank Felipe Morales for technical support and stimulating discussions.
DA and OS acknowledge support from the DFG SPP 1840 ``Quantum Dynamics in Tailored Intense Fields'' and DFG grant SM 292/5-1. 
DA acknowledges support from the Royal Society University Research Fellowship URF$\backslash$R1$\backslash$201333.

\section*{Competing Interests}
The authors declare that they have no competing financial interests.
The data that support the plots within this paper and other findings of this study are available from the corresponding authors upon reasonable request. 
Correspondence should be addressed to david.ayuso@mbi-berlin.de and olga.smirnova@mbi-berlin.de. 

\section*{References}
\bibliography{Bibliography}

\end{document}